\documentclass[aps,pre,twocolumn]{revtex4-2}

\usepackage{graphicx}
\usepackage{amsmath}
\usepackage{bm}

\begin{document}

\title{Spontaneous Cholesteric Phase in Ferroelectric Nematic Liquid Crystals:  Preference for Integer Number of Pitches}
\author{Lincoln Paik}
\author{Jonathan V. Selinger}
\affiliation{Department of Physics, Advanced Materials and Liquid Crystal Institute, Kent State University, Kent, Ohio 44242, USA}
\date{July 3, 2025}
\begin{abstract}
In a ferroelectric nematic liquid crystal, the electrostatic interaction can induce a spontaneous cholesteric helix, even if the material is not chiral.  If the liquid-crystal cell is infinitely thick, then the predicted pitch depends continuously on material parameters.  Here, we consider how the prediction must be modified in a cell of finite thickness.  If the Debye screening length is large enough, we find that the free energy has multiple minima.  In these minima, the cholesteric pitch is locked to the cell thickness, so that the cell contains an integer number of pitches.  However, if the screening length is smaller, then the cholesteric pitch can vary continuously.
\end{abstract}

\maketitle

\section{Introduction}
Liquid crystals can exhibit phases with different types of orientational order.  The conventional nematic phase has molecules aligned in both directions along an axis, so that it has twofold symmetry, with no polarity.  By contrast, the ferroelectric nematic phase has molecules aligned in a single direction, so that it has an electrostatic polarization.  The ferroelectric nematic phase was proposed theoretically by Born more than 100 years ago~\cite{Born1916}, and it has only recently been discovered experimentally in newly synthesized materials~\cite{Chen2020,Lavrentovich2020,Sebastian2022,Mandle2022}.  Since this discovery, it has become a major subject of experimental and theoretical research~\cite{Zou2024,MedleRupnik2025,Basnet2025,Mathe2025,Ma2025}.

In 1975, long before the experimental discovery of ferroelectric nematic liquid crystals, Khachaturyan made a remarkable prediction about their configuration~\cite{Khachaturyan1975}.  He noted that a uniform alignment of the electrostatic polarization would have a very high electrostatic energy, and this energy could be reduced by twisting the polarization field into a helix.  For that reason, he argued that the ground state could actually be a helix, as shown in Fig.~1.  This configuration would be similar to a cholesteric liquid crystal, but it would not be induced by molecular chirality.  Rather, it could be randomly right- or left-handed, as an example of spontaneous symmetry breaking.  In 2024, Kumari \emph{et al.}\ reported experimental observations of this type of helix in a ferroelectic nematic liquid crystal~\cite{Kumari2024}.  They also pointed out a typographical error in Khachaturyan's paper, with the wrong dimensions in the prediction for the cholesteric pitch.  In a recent paper, we repeated Khachaturyan's calculation using a different theoretical method and corrected the pitch prediction~\cite{Paik2025}.  Lavrentovich \emph{et al.}\ have done further experiments and calculations on twisted configurations in cells with specified surface anchoring~\cite{Lavrentovich2025}.

\begin{figure}
\centering
\includegraphics[width=0.5\columnwidth]{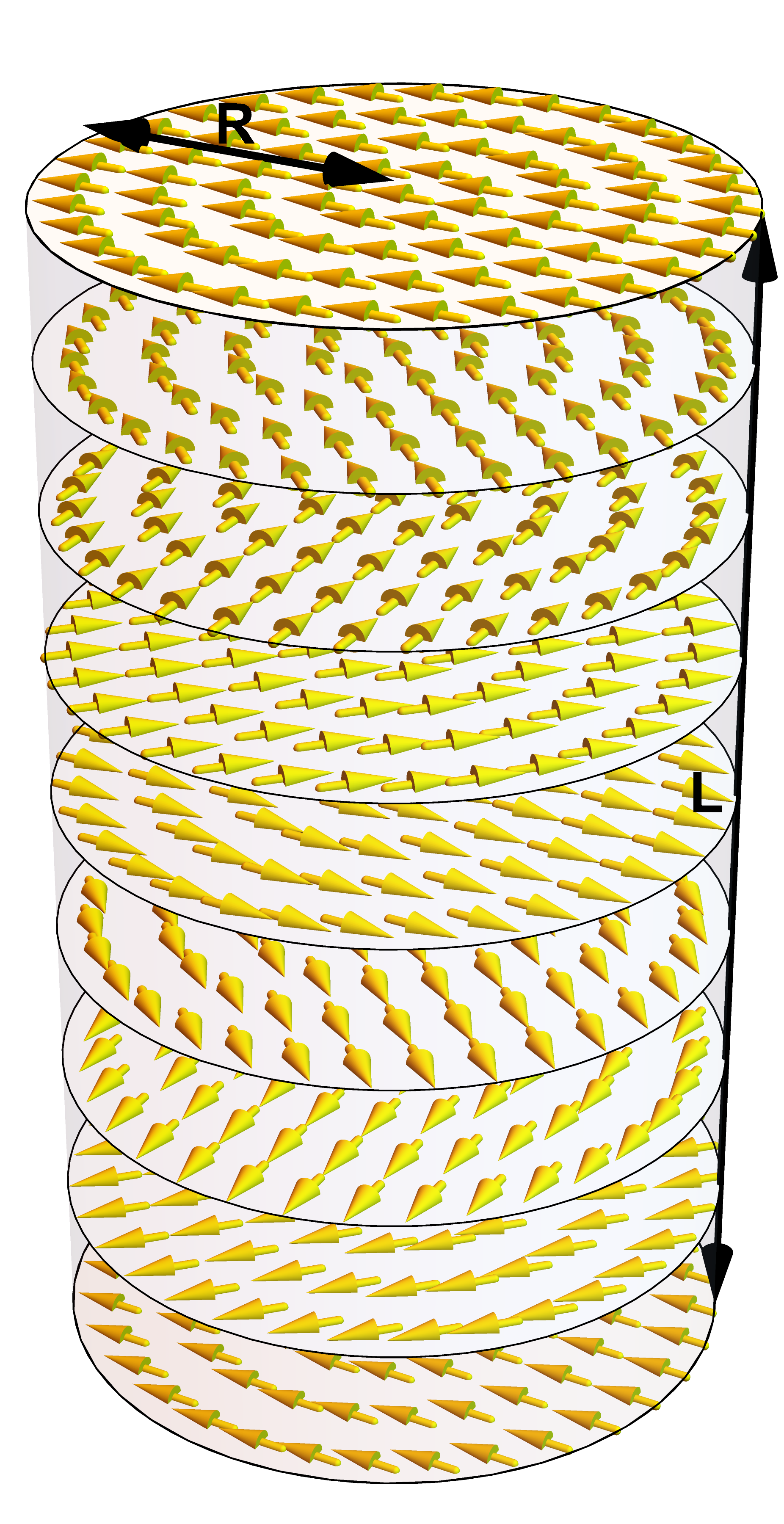}
\caption{Configuration of the polarization field $\bm{P}(\bm{r})$ inside a cylindrical domain of radius $R$ and height $L$.  In this example, the pitch of the spontaneous helix is equal to $L$.}  
\end{figure}

In the Khachaturyan theory, the pitch of the spontaneous helix is determined by competition between the electrostatic energy (which favors a short pitch) and the twist elastic free energy (which favors a long pitch).  As a result, the predicted pitch depends continuously on material parameters.  This prediction applies in the limit of infinite cell thickness, so that there is no constraint on the number of pitches across the thickness.  However, if the cell has a finite thickness, then the considerations become more complex.  In a finite cell, the electrostatic energy is lowest if the number of pitches across the cell is an integer, and any fractional pitch costs extra energy.  For that reason, the pitch might be effectively constrained, so that the cell thickness is an integer multiple of the pitch.  This constraint would appear as a spontaneous boundary condition---not caused by surface anchoring, but rather by the bulk electrostatic energy.

The purpose of this paper is to determine when the pitch is constrained by the cell thickness, as opposed to varying continuously with material parameters.  For that purpose, we generalize our previous calculation~\cite{Paik2025} to a cell with finite thickness.  In this finite cell, we calculate the total of electrostatic and elastic free energy as a function of the wavevector of the spontaneous helix.  Our results depend on the Debye screening length
\begin{equation}
\Lambda=\sqrt{\frac{\epsilon k_B T}{e^2 n}},
\label{Debye}
\end{equation}
where $\epsilon$ is the dielectric constant, $k_B T$ the thermal energy, $e$ the electron charge, and $n$ the concentration of free ions in the material.  If $\Lambda$ is sufficiently long, the free energy plot shows a series of multiple minima, corresponding to an integer number of pitches across the cell.  However, if $\Lambda$ is shorter, then the plot shows only a single minimum, corresponding to a pitch that varies continuously.  Hence, both types of behavior can be expected in experimental systems.

\section{Theory}

We consider a liquid crystal in a cylindrical geometry with height $L$ and radius $R$, as shown in Fig.~1.  This geometry represents a typical liquid crystal cell, in which $L$ corresponds to the cell thickness and $R$ to the lateral domain size.  We assume that the director field has the helical configuration $\hat{\bm{n}}(\bm{r})=(\cos q z,\sin q z,0)$, with the pitch of $2\pi/q$.  Also, we assume that the polarization field is $\bm{P}(\bm{r})=P_0\hat{\bm{n}}(\bm{r})$, where the magnitude $P_0$ is constant.

The total free energy of the liquid crystal $F=F_\text{Frank}+F_\text{elec}$ includes an elastic component and an electrostatic component.  The first term is the Frank free energy
\begin{align}
F_\text{Frank}&=\int d^3 r \biggl[\frac{1}{2}K_{11}(\bm{\nabla}\cdot\hat{\bm{n}})^2
+\frac{1}{2}K_{22}[\hat{\bm{n}}\cdot(\bm{\nabla}\times\hat{\bm{n}})]^2\nonumber\\
&\qquad\qquad+\frac{1}{2}K_{33}|\hat{\bm{n}}\times(\bm{\nabla}\times\hat{\bm{n}})|^2\biggr],
\end{align}
where $K_{11}$, $K_{22}$, and $K_{33}$ are the Frank elastic constants for splay, twist, and bend, respectively.  For the helical director configuration, the splay and bend are both zero, and the Frank free energy becomes simply
\begin{equation}
F_\text{Frank}=(\pi R^2 L)\left(\frac{1}{2}K_{22}q^2\right).
\label{Frank}
\end{equation}

The second term is the screened electrostatic interaction
\begin{equation}
F_\text{elec}=\frac{1}{2}\int d^3 r_1 d^3 r_2 \frac{\rho(\bm{r}_1)\rho(\bm{r}_2)e^{-|\bm{r}_{12}|/\Lambda}}{4\pi\epsilon|\bm{r}_{12}|},
\end{equation}
which couples the electric charge density $\rho$ at positions $\bm{r}_1$ and $\bm{r}_2$, with $\bm{r}_{12}=\bm{r}_2-\bm{r}_1$.  The parameter $\Lambda$ is the Debye screening length, with $\Lambda\to\infty$ corresponding to an unscreened Coulomb interaction.  The parameter $\epsilon$ is the dielectric constant; for simplicity we assume that it is isotropic.  The prefactor of $\frac{1}{2}$ prevents double counting.

In our previous paper~\cite{Paik2025}, we considered two other terms in the free energy that are not relevant here.  One of those terms is the flexoelectric energy, which is zero because the helical director configuration does not have any splay or bend.  The other is the Landau free energy as a power series in the polarization, which is constant because the polarization magnitude $P_0$ is assumed to be constant.  Any gradient term in the Landau free energy can be absorbed into an effective Frank constant $K_{22}$.

The next step is to calculate the electrostatic interaction $F_\text{elec}$ for the helical configuration in the cylindrical geometry with finite $L$, without assuming any relationship between $q$ and $L$.  In general, the electric charge density is the sum of bound and surface charge densities, $\rho(\bm{r})=\rho_\text{bound}+\rho_\text{surface}$.  For the helical configuration, the bound charge density is $\rho_\text{bound}(\bm{r})=-\bm{\nabla}\cdot\bm{P}=0$ because there is no splay.  The surface charge density per volume is $\rho_\text{surface}(\bm{r})=(\hat{\bm{N}}\cdot\bm{P})\delta(\hat{\bm{N}}\cdot\bm{r}-R)$, where $\hat{\bm{N}}$ is the local unit vector normal to the surface.  Working in cylindrical coordinates $(r,\phi,z)$, this normal vector is just $\hat{\bm{N}}=\hat{\bm{r}}$.  Hence, the surface charge density per volume is
\begin{equation}
\rho_\text{surface}(\bm{r})=P_0 \cos(q z - \phi)\delta(r-R).
\end{equation}
The electrostatic energy is then
\begin{align}
F_\text{elec}&=\frac{P_0^2 R^2}{2}\int d\phi_1 dz_1 d\phi_2 dz_2
\frac{e^{-|\bm{r}_{12}|/\Lambda}}{4\pi\epsilon |\bm{r}_{12}|}\times\nonumber\\
&\qquad\qquad\qquad\cos(q z_1 - \phi_1)\cos(q z_2 - \phi_2),
\end{align}
with
\begin{equation}
|\bm{r}_{12}|=\sqrt{4R^2\sin^2[\textstyle{\frac{1}{2}}(\phi_1-\phi_2)]+(z_1 -z_2)^2}.
\end{equation}
We change variables to $\bar{\phi}=\frac{1}{2}(\phi_1+\phi_2)$ and $\delta\phi=\phi_1-\phi_2$, and integrate over $\bar{\phi}$, to obtain
\begin{align}
F_\text{elec}&=\frac{P_0^2 R^2}{8\epsilon}\int dz_1 dz_2 d(\delta\phi)
\frac{e^{-|\bm{r}_{12}|/\Lambda}}{|\bm{r}_{12}|}\cos(q z_1 - q z_2 - \delta\phi)\nonumber\\
&=\int dz_1 dz_2 f(|z_1 - z_2|) \cos(q z_1 - q z_2),
\end{align}
where
\begin{equation}
f(|z_1-z_2|)=\frac{P_0^2 R^2}{8\epsilon}
\int d(\delta\phi)\frac{e^{-|\bm{r}_{12}|/\Lambda}}{|\bm{r}_{12}|}\cos\delta\phi.
\end{equation}
Following the same method as in the Appendix of Ref.~\cite{Paik2025}, we can calculate the Fourier transform of $f(z)$ as
\begin{equation}
\tilde{f}(k)=\frac{1}{2\pi}\int_{-\infty}^\infty dz e^{-i k z}f(z)
=\frac{P_0^2 R}{8\epsilon\sqrt{k^2+\Lambda^{-2}}}.
\end{equation}
The inverse Fourier transform is then
\begin{equation}
f(z)=\int_{-\infty}^\infty dk e^{i k z}\tilde{f}(k)
=\frac{P_0^2 R}{4\epsilon}K_0\left(\frac{|z|}{\Lambda}\right),
\end{equation}
where $K_0$ is a modified Bessel function of the second kind.  Hence, the electrostatic energy becomes
\begin{equation}
F_\text{elec}=\frac{P_0^2 R}{4\epsilon}\int dz_1 dz_2 K_0\left(\frac{|z_1-z_2|}{\Lambda}\right) \cos{(q z_1 -q z_2)}.
\label{integralversion1}
\end{equation}
Alternatively, by changing variables to $\bar{z}=\frac{1}{2}(z_1+z_2)$ and $\delta z=z_1-z_2$, we can express it as
\begin{equation}
F_\text{elec}=\frac{P_0^2 R}{4\epsilon}\int d\bar{z} d(\delta z)
K_0\left(\frac{|\delta z|}{\Lambda}\right) \cos(q\delta z).
\label{integralversion2}
\end{equation}

In these expressions, the limits of integration are finite because the cell thickness is finite.  In Eq.~(\ref{integralversion1}), we must integrate over the domain
\begin{equation}
|z_1|\leq\frac{L}{2}\text{ and }|z_2|\leq\frac{L}{2}.
\end{equation}
Likewise, in Eq.~(\ref{integralversion2}), we must integrate over the domain
\begin{subequations}
\begin{equation}
|\bar{z}|\leq\frac{L}{2}\text{ and }|\delta z|\leq L-2|\bar{z}|,
\end{equation}
or equivalently
\begin{equation}
|\delta z|\leq L\text{ and }|\bar{z}|\leq\frac{L}{2}-\frac{|\delta z|}{2}.
\end{equation}
\end{subequations}
Using the latter set of limits, Eq.~(\ref{integralversion2}) becomes
\begin{align}
F_\text{elec}&=\frac{P_0^2 R}{4\epsilon}\int_{-L}^L d(\delta z) \int_{-(L-|\delta z|)/2}^{(L-|\delta z|)/2} d\bar{z} 
K_0\left(\frac{|\delta z|}{\Lambda}\right)\cos(q\delta z)\nonumber\\
&=\frac{P_0^2 R}{2\epsilon}\int_0^L d(\delta z) (L-\delta z)
K_0\left(\frac{\delta z}{\Lambda}\right) \cos(q\delta z).
\label{felecintegral}
\end{align}
We cannot evaluate this integral exactly, but we can estimate it in two limiting cases:

\paragraph{Cell thickness large compared with Debye screening length}  For $L\gg\Lambda$, we can extend the upper limit of the integral to infinity, because the Bessel function $K_0(\delta z/\Lambda)$ is exponentially small in the range from $\delta z=L$ to infinity.  The electrostatic free energy then becomes
\begin{align}
F_\text{elec}&=
\frac{P_0^2 R[\pi L(q^2+\Lambda^{-2})-2\sqrt{q^2+\Lambda^{-2}}+2q\sinh^{-1}q\Lambda]}{4\epsilon(q^2+\Lambda^{-2})^{3/2}}.
\label{FelecfinitebigL}
\end{align}
In the limit of $L\to\infty$, this result reduces to
\begin{equation}
F_\text{elec}=(\pi R^2 L)\left(\frac{P_0^2}{4R\epsilon\sqrt{q^2+\Lambda^{-2}}}\right),
\label{Felecinfinite}
\end{equation}
which is the electrostatic energy calculated in Ref.~\cite{Paik2025}.

\paragraph{Cell thickness small compared with Debye screening length} For $L\ll\Lambda$, we can replace the Bessel function by its series expansion $K_0(\delta z/\Lambda)\approx\log(2\Lambda/\delta z)-\gamma$, where $\gamma\approx0.577$ is Euler's constant.  The integral of Eq.~(\ref{felecintegral}) can then be evaluated as
\begin{align}
F_\text{elec}&=\frac{P_0^2 R}{2 \epsilon q^2}
\biggl[-1 - \mathrm{Ci}(L q) + \cos(L q) \left(1+ \gamma + \log\frac{L}{2\Lambda}\right)\nonumber\\
&\qquad\qquad+\log(2 q\Lambda)+L q \mathrm{Si}(L q)\biggr],
\label{FelecfinitesmallL}
\end{align}
where $\mathrm{Si}(x)=\int_{0}^{x}\sin t dt/t$ and  $\mathrm{Ci}(x)=-\int_{x}^{\infty}\cos t dt/t$ are the sine integral and cosine integral functions, respectively.

\section{Analysis}

We must now minimize the total free energy $F=F_\text{Frank}+F_\text{elec}$ over the helical wavevector $q$.  

To begin, let us review the infinite system $L\to\infty$, which was considered by Khachaturyan~\cite{Khachaturyan1975} and our previous paper~\cite{Paik2025}.  In this case, the free energy combines the Frank free energy of Eq.~(\ref{Frank}) with the electrostatic free energy of Eq.~(\ref{Felecinfinite}).  The minimization result depends on the Debye screening length $\Lambda$ compared with the critical value $\Lambda_c = (4 K_{22}R\epsilon/P_0^2)^{1/3}$.  If $\Lambda>\Lambda_c$, the minimum occurs at $q=\sqrt{\Lambda_c ^{-2} -\Lambda ^{-2}}$.  Hence, the liquid crystal forms a helix, with a pitch that depends continuously on material parameters.  However, if $\Lambda<\Lambda_c$, the minimum occurs at $q=0$, so that the liquid crystal configuration is uniform.

In the experiment of Kumari \emph{et al.}~\cite{Kumari2024}, typical parameters are $\epsilon= 10^{-9}$ C$^2$/(N m$^2$), $K_{22}= 5$ pN, $R = 50$ $\mu$m, and $P_0 = 4.4 \times 10^{-2}$ C/m$^2$.  With those parameters, the critical Debye screening length is $\Lambda_c = 80$ nm.  From Eq.~(\ref{Debye}), at room temperature, the corresponding critical ion concentration is $n_c = 2.5\times 10^{22}$ m$^{-3}$.

\begin{figure}
\centering
\includegraphics[width=241pt]{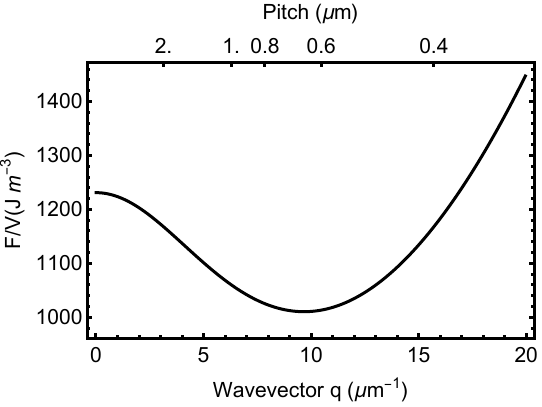}
\caption{Total free energy per volume for an infinite system, as a function of wavevector $q$ or pitch $2\pi/q$.  Parameters are $\epsilon= 10^{-9}$ C$^2$/(N m$^2$), $K_{22}= 5$ pN, $R = 50$ $\mu$m, and $P_0 = 4.4 \times 10^{-2}$ C/m$^2$.  In this example, the ion concentration is $n = 1 \times 10^{22}$ m$^{-3}$, and hence the Debye screening length is $\Lambda=127$ nm.}
\end{figure}

For a specific example, Fig.~2 shows the free energy per volume $F/(\pi R^2 L)$ as a function of wavevector $q$ for the infinite system with ion concentration $n = 1\times 10^{22}$ m$^{-3}$, or Debye screening length $\Lambda=127$ nm.  Because $\Lambda>\Lambda_c$, the plot has a minimum at finite $q=9.7$ $\mu$m$^{-1}$, corresponding to a pitch of $2\pi/q=0.65$ $\mu$m.  This pitch varies continuously with the parameters, and is not locked at any specific value.

\begin{figure}
\centering
\includegraphics[width=241pt]{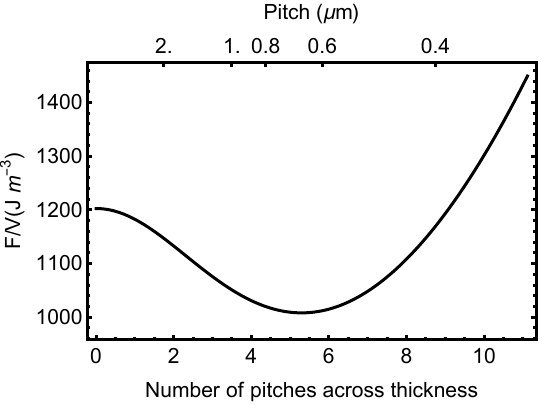}
\caption{Total free energy per volume for a cell of thickness $L=3.5$ $\mu$m, as a function of the pitch $2\pi/q$, or equivalently as a function of the number of pitches across the thickness $N_\text{pitch}=Lq/(2\pi)$.  Numerical parameters are the same as in Fig.~2.} 
\end{figure}

Now suppose the cell thickness $L$ is finite but larger than the Debye screening length $\Lambda$.  In this case, the free energy combines the Frank free energy of Eq.~(\ref{Frank}) with the electrostatic free energy of Eq.~(\ref{FelecfinitebigL}).  As an example, Fig.~3 shows the free energy per volume for $L=3.5$ $\mu$m, still with $\Lambda=127$ nm.  In this graph, the bottom axis is changed from wavevector $q$ to the number of pitches across the thickness, which is $N_\text{pitch}=L q/(2\pi)$.  The plot is quite similar to Fig.~2.  It has a minimum at finite $q=9.5$ $\mu$m$^{-1}$, corresponding to pitch $2\pi/q=0.66$ $\mu$m and to $N_\text{pitch}=5.3$.  As in the previous case, these quantities vary continuously as functions of the material parameters; $N_\text{pitch}$ is not locked at any integer or half-integer value.

\begin{figure*}
\centering
(a)\includegraphics[width=241pt]{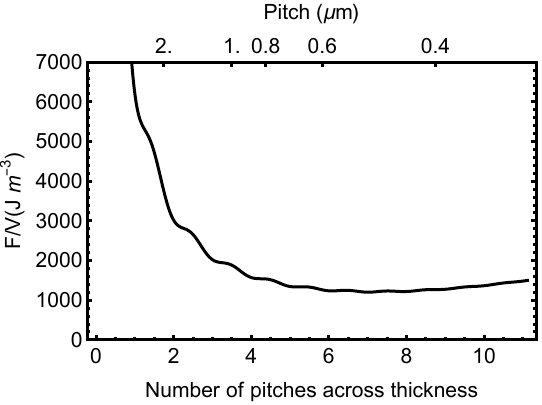}
(b)\includegraphics[width=241pt]{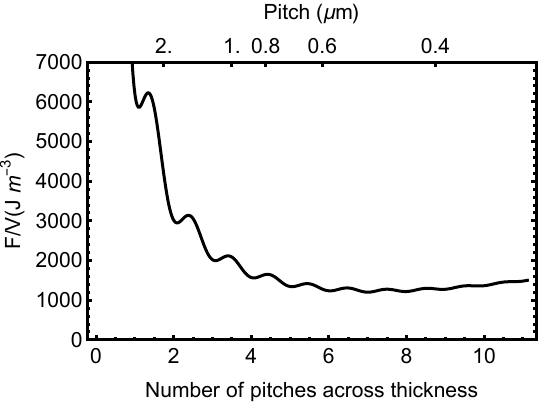}
(c)\includegraphics[width=241pt]{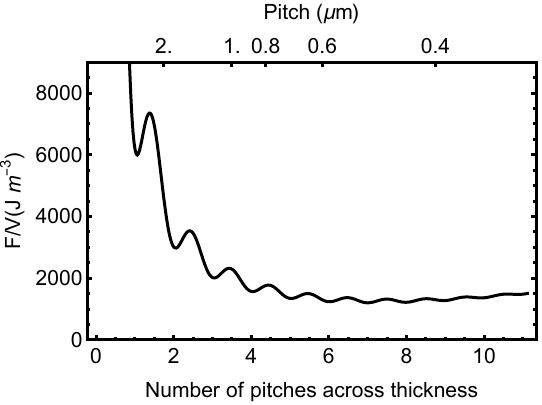}
(d)\includegraphics[width=241pt]{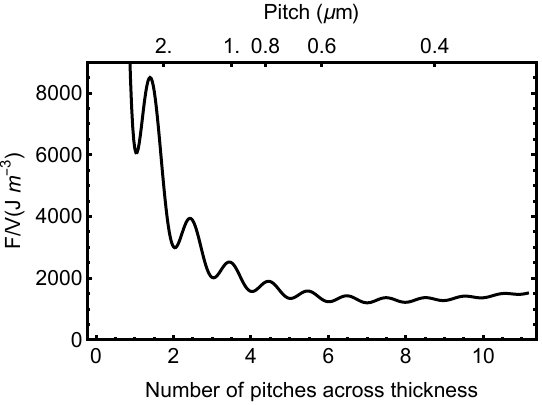}
\caption{Total free energy per volume for a cell of thickness $L=3.5$ $\mu$m, as a function of the pitch $2\pi/q$, or equivalently as a function of the number of pitches across the thickness $N_\text{pitch}=Lq/(2\pi)$.  The graphs show a series of ion concentrations, and hence a series of Debye screening lengths:
(a) $n=10^{18}$ m$^{-3}$, $\Lambda = 12.7$ $\mu$m.
(b) $n=10^{16}$ m$^{-3}$, $\Lambda = 127$ $\mu$m.
(c) $n=10^{14}$ m$^{-3}$, $\Lambda = 1.27$ mm.
(d) $n=10^{12}$ m$^{-3}$, $\Lambda = 12.7$ mm.}
\end{figure*} 

For comparison, suppose the Debye screening length $\Lambda$ is larger than the cell thickness $L$, so that the free energy combines the Frank free energy of Eq.~(\ref{Frank}) with the electrostatic free energy of Eq.~(\ref{FelecfinitesmallL}).  Figure~4 shows a series of free energy plots with the ion concentration $n$ decreasing from $10^{18}$ to $10^{12}$ m$^{-3}$, and hence $\Lambda$ increasing from 12.7 $\mu$m to 12.7 mm, all with $L=3.5$ $\mu$m.  In this series of plots, we can see that the free energy develops multiple minima, which occur at integer values of $N_\text{pitch}=L q/(2\pi)$.  As $\Lambda$ increases, these multiple minima become deeper.  The free energy minimum occurs at a value of $q$ that is similar to the value in the infinite system, but it is modified so that $N_\text{pitch}$ is rounded off to the nearest integer.  Hence, the pitch does not vary continuously with material parameters, but rather it is locked to an integer relationship with the cell thickness.

Mathematically, the multiple minima can be understood from the electrostatic energy expression in Eq.~(\ref{FelecfinitesmallL}).  They arise from the term proportional to $\cos(Lq)$, which has a coefficient of $1+\gamma+\log[L/(2\Lambda)]$.  This coefficient is large and negative if $\Lambda\gg L$.  In that case, it favors $L q$ to be an integer multiple of $2\pi$, and hence it favors $N_\text{pitch}$ to be an integer.  (In the opposite limit of $\Lambda\ll L$, the electrostatic energy is given by Eq.~(\ref{FelecfinitebigL}) rather than (\ref{FelecfinitesmallL}), and hence this coefficient does not matter.)

Physically, the multiple minima occur because the electrostatic polarization averages to zero whenever the cell thickness contains an integer number of helical pitches.  This cancellation greatly reduces the electrostatic energy.  In the original paper~\cite{Khachaturyan1975}, when Khachaturyan considered the unscreened electrostatic interaction with $\Lambda\to\infty$, he argued that the ground state must have a polarization that integrates to zero $\int d^3 r \bm{P}(\bm{r})=0$.  Here, for $\Lambda$ large but finite, we see that the free energy has a finite preference for the polarization to integrate to zero.  This preference leads to multiple minima at integer values of $N_\text{pitch}$ provided that $\Lambda\gg L$.

\section{Discussion}

The behavior discussed in this paper can be compared with classic experiments on cholesteric liquid crystals in Grandjean-Cano wedge cells~\cite{Grandjean1921,Smalyukh2002}.  In the classic experiment, a cholesteric liquid crystal is confined to a wedge-shaped cell with strong anchoring on the top and bottom surfaces.  The strong anchoring requires the cholesteric helix to have an integer or half-integer number of pitches across the cell thickness.  Along the wedge, as the cell thickness changes, the number of pitches must jump from one allowed value to another.  Hence, the liquid crystal forms a series of dislocations in the layer structure.  These dislocations are observed experimentally as a series of parallel lines in polarized optical microscopy.

Here, we see a similar effect for a completely different reason.  When a ferroelectric nematic liquid crystal forms a cholesteric helix, it will have an integer number of pitches across the cell thickness, provided that the Debye screening length $\Lambda$ is long enough.  If the liquid crystal is placed into a wedge geometry, it should also form stripes with different integer numbers of pitches, separated by dislocations in the layer structure.  This effect is not caused by surface anchoring, but rather by the bulk electrostatic energy.  Indeed, it would be suppressed by strong surface anchoring, which would control the number of pitches in a different way.  The effect could only be seen in liquid crystal cells with free surfaces.  It provides an example of the rich range of behavior that arises from the combination of elasticity and electrostatics in ferroelectric nematic liquid crystals.

\acknowledgments
We thank P. Kumari, M.~O. Lavrentovich, and O.~D. Lavrentovich for helpful discussions.  This work was supported by National Science Foundation Grant DMR-1409658.

\bibliography{version2}

\end{document}